\documentclass[a4paper,11pt]{article}
\pdfoutput=1 

\usepackage{jinstpub} 
\usepackage{siunitx}
\usepackage{xargs}   
\usepackage{soul}
\usepackage{siunitx}
\usepackage{textcomp}
\usepackage{todonotes}

\usepackage{tabularx}
\usepackage{multicol}
\usepackage{subcaption}

\usepackage{tablefootnote}

\DeclareSIUnit[number-unit-product = ]\percent{\char`\%}

\title{\boldmath The upgraded low-background germanium  counting facility Gator for high-sensitivity $\gamma$-ray spectrometry
}

\author{Gabriela R. Araujo,}
\author{Laura Baudis,}
\author{Yanina Biondi,}
\author{Alexander Bismark,}
\author{Michelle Galloway}

\affiliation{Department of Physics, University of Zurich, Winterthurerstrasse 190, CH-8057 Zurich, Switzerland}

\emailAdd{gabriela.araujo@physik.uzh.ch}
\emailAdd{alexander.bismark@physik.uzh.ch}

\abstract{We describe the upgrade and performance of the high-purity germanium counting facility Gator, which is dedicated to low-background  $\gamma$-ray spectrometry. Gator is operated at the Gran Sasso Underground Laboratory in Italy, at an average depth of 3600 meter water equivalent, and employed for material screening and selection in ultra-low background, rare-event search experiments in astroparticle physics. The detector is equipped with a passive shield made of layers of copper, lead and polyethylene, and the sample cavity is purged with gaseous nitrogen maintained at positive pressure for radon suppression. After upgrading its enclosure, the background rate is \SI{82.0(7)}{counts/(kg\cdot day)} in the energy region \SIrange{100}{2700}{keV}, a \SI{20}{\percent} reduction compared to the previously reported rate. We show the stability of various operation parameters as a function of time. We also summarize the sample analysis procedure, and demonstrate Gator's sensitivity by examining one material sample, a candidate photosensor for the DARWIN experiment.}

\begin{document}
\maketitle
\flushbottom

\section{Introduction}
\label{sec:intro}
Gamma-ray spectrometers based on ultra-low background, high-purity germanium (HPGe) detectors are routinely employed to screen and select detector materials for rare-event searches~\cite{Heusser:1995wd}.
Germanium spectrometry offers a non-destructive screening method and provides information on material radioactivity levels in a single energy spectrum with excellent resolution for $\gamma$-ray lines emitted by primordial ($^{238}$U, $^{232}$Th, $^{40}$K), cosmogenic (e.g.~$^{54}$Mn, $^{46}$Sc, $^{60}$Co), and anthropogenic (e.g.~$^{137}$Cs, $^{110\mathrm{m}}$Ag) isotopes. Sensitivities for HPGe detectors located in underground laboratories are at the level of $10-\SI{100}{\micro Bq/kg}$ for large masses (tens of kg) and long counting times (1-3 months)~\cite{Heusser:2006495,Baudis:2011am,Heusser:2015ifa,Loaiza:2015dma, Sivers:2016yvm, Garcia:2022jdt}. 

The Gator facility, located at the Laboratori Nazionali del Gran Sasso (LNGS) of INFN in Italy and described in detail in~\cite{Baudis:2011am}, has been employed to select materials for the dark matter search experiments XENON100~\cite{Aprile:2011ru}, XENON1T~\cite{Aprile:2017ilq} and XENONnT~\cite{XENON:2021mrg}, as well as for the neutrinoless double beta decay experiments GERDA~\cite{Agostini:2017hit} and LEGEND-200. Radioassay of potential materials for the DARWIN~\cite{Aalbers:2016jon} experiment is ongoing and will also be performed for LEGEND-1000~\cite{Abgrall:2017syy} in the near future. We have upgraded the facility in order to decrease the overall background level as well as to facilitate the sample handling process. The aim of this paper is to describe the facility modifications and their impact on the background rate with respect to~\cite{Baudis:2011am} as well as provide an update on the detector operation and calibration. We also present the sample analysis procedure, illustrated with a sample from the DARWIN material screening campaign. This article is organised as follows: In \autoref{sec:upgrade} we describe the improvements to the system in detail. In \autoref{sec:operation} we describe the detector operation and general performance, and we discuss the new background level in \autoref{sec:background}. In \autoref{sec:samples} we present the analysis procedure along with results for a DARWIN sample which was measured after the upgrade. We provide a summary and outlook in \autoref{sec:summary}.

\section{Description of the detector and its upgrade}
\label{sec:upgrade}
Gator deploys a p-type coaxial HPGe detector with a \SI{2.2}{kg} sensitive mass and a relative efficiency of \SI{100.5}{\percent}\footnote{Detection efficiency given for the \SI{1.33}{MeV} $\gamma$-ray from $^{60}$Co and defined relative to a 3-inch × 3-inch NaI(Tl) crystal placed at a distance of \SI{25}{cm} from the source ~\citep{Knoll}.}~\cite{Baudis:2011am}. The HPGe crystal is placed in an ultra-low background, oxygen-free copper cryostat and surrounded by several layers of shielding material. The innermost \SIrange{5}{7}{cm}-thick shield is made of oxygen-free high-conductivity copper, which is surrounded by 5-cm inner and 15-cm thick outer layers of lead with activities of \SI{3}{Bq/kg} and \SI{75}{Bq/kg}, respectively. The outermost, 5-cm thick shield is made of borated polyethylene in order to reduce the ambient neutron flux. The sample cavity has inner dimensions of 25$\times$25$\times$\SI{33}{cm^3} and is continuously purged with gaseous nitrogen ($\mathrm{GN}_2$) at positive pressure to displace environmental radon. The detector is operated at cryogenic temperatures, with cooling provided by a copper coldfinger immersed in liquid nitrogen ($\mathrm{LN}_2$). The detector design and operation follow all the relevant recommendations of ISO 20042 on technical requirements for $\gamma$-spectrometry methods~\citep{ISO20042}. For details, we refer to~\cite{Baudis:2011am}.

During the first ten-year period of operation at LNGS (2007-2017), the shielding structure was surrounded by an aluminum housing with an acrylic glove box placed above it for sample handling, as shown in~\cite{Baudis:2011am}. An airlock system attached to the glove box allowed for pre-purging with $\mathrm{GN}_2$ as well as storage of samples within a nitrogen atmosphere prior to insertion into the sample cavity. The latter allowed for the decay of $^{222}$Rn ($\mathrm{T}_{1/2}=\SI{3.82}{days}$), which may be introduced during sample exchange. In this initial design, both the geometry of the airlock and the \mbox{acrylic-to-aluminum} interface of the glove box made it difficult to optimally purge against radon, as well as to access the cavity for sample loading. Motivated by these problems and by the requirement to clean the cavity and shield after years of operation, the enclosure and airlock were redesigned, fabricated, and installed.

\begin{figure}
\centering
\includegraphics[width=0.9\textwidth]{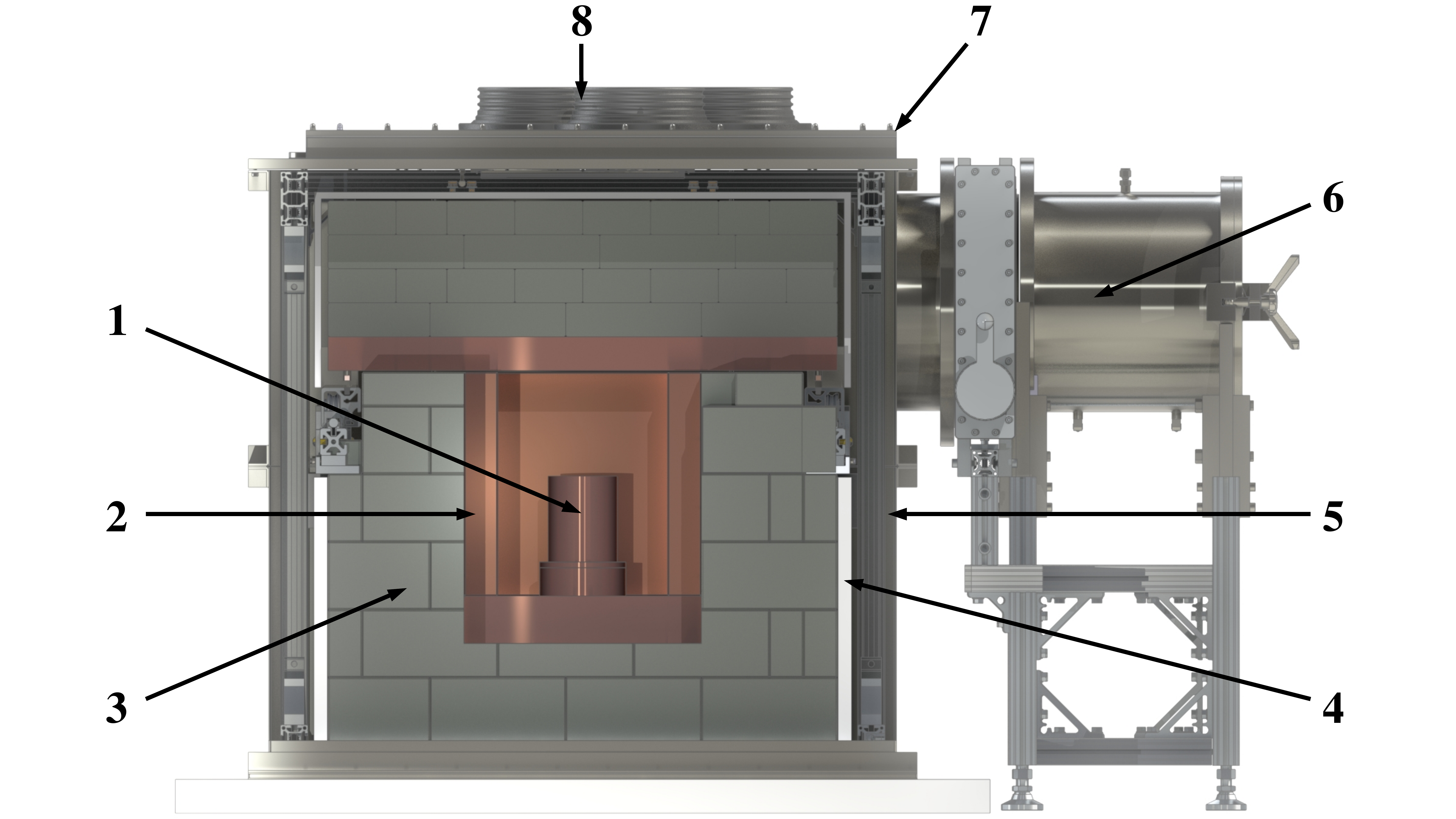}
\caption{Section view of the Gator setup after the upgrade. The HPGe crystal is housed inside a copper cryostat (1) within the sample cavity, which is purged with ($\mathrm{GN}_2$). The cavity is formed by surrounding layers of low-background copper (2), lead (3) and polyethylene (4). These layers were disassembled, cleaned and rebuilt within the new enclosure (5). The load-lock chamber (6) allows for pre-purging of samples. The acrylic plate (7) and glove ports (8) for sample viewing and handling are also shown.}
\label{fig:gator_setup}
\end{figure}

The new enclosure, made of stainless steel, is shown schematically in \autoref{fig:gator_setup}. The base of the enclosure consists of two vertically stacked, rectangular pieces and a flat top plate, with each layer sealed by a continuous rubber gasket. In place of the acrylic glove box is a flat acrylic plate mounted onto the top plate and sealed by a rubber gasket, which houses several glove ports. These design aspects significantly improved the hermeticity of the system, thus prohibiting the entry of radon via air leaks. The acrylic plate accommodates several glove ports that allow for top-loading of samples, facilitating access to the entire volume of the cavity. A removable flange housing the electrical connections is also installed onto the top plate and sealed with a rubber gasket. This flange additionally contains ports that allow for regular nitrogen fills, as the ($\mathrm{LN}_2$) dewar connected to the detector coldfinger is within the enclosure.

The acrylic airlock was replaced by a stainless steel, circular load-lock chamber (shown to the right of the enclosure, \autoref{fig:gator_setup}), where samples are pre-purged with $\mathrm{GN}_2$ before being placed inside the cavity. Once purged, the sliding shield on the top of the cavity is opened, followed by the gate valve that separates the two volumes. The sample is then directly loaded into the cavity from the inside of the load-lock chamber. 

To prepare for installation, the old enclosure was removed and the lead and copper shielding disassembled. The innermost shielding layers were cleaned in an ultrasonic ethanol bath in order to remove $^{40}$K, oils, and particulates from surfaces. The structure was then rebuilt within the base of the new enclosure. Afterwards the $\mathrm{GN}_2$ flow was optimized for the new configuration. The improved hermiticity allowed for a 2/3 reduction in flow at the inlet while maintaining the same outlet flow as in the previous enclosure. Further optimisations of the system include the implementation of hermetic seals on all of the electrical connections that pass through the enclosure and additional shielding on detector cabling to reduce noise. The integration of a normally-closed switch on the power line was also added to prevent the automatic ramp-up of the high voltage module following a power outage.

\section{Detector operation and performance}
\label{sec:operation}

The HPGe detector is operated at full depletion under a bias voltage of \SI{4.8}{kV}. The applied voltage, along with other parameters that are critical indicators of detector performance and possible failure (leakage current and $\mathrm{LN}_2$ level) are monitored: Their values are written into a database every two minutes. If exceeding a specified range, alarms are triggered and sent via email and SMS. We also monitor the room temperature, which could affect the gain of the amplifier and thus the calibration of the detector~\citep{ISO20042}. 

\begin{figure}
\centering
\includegraphics[width=1\textwidth]{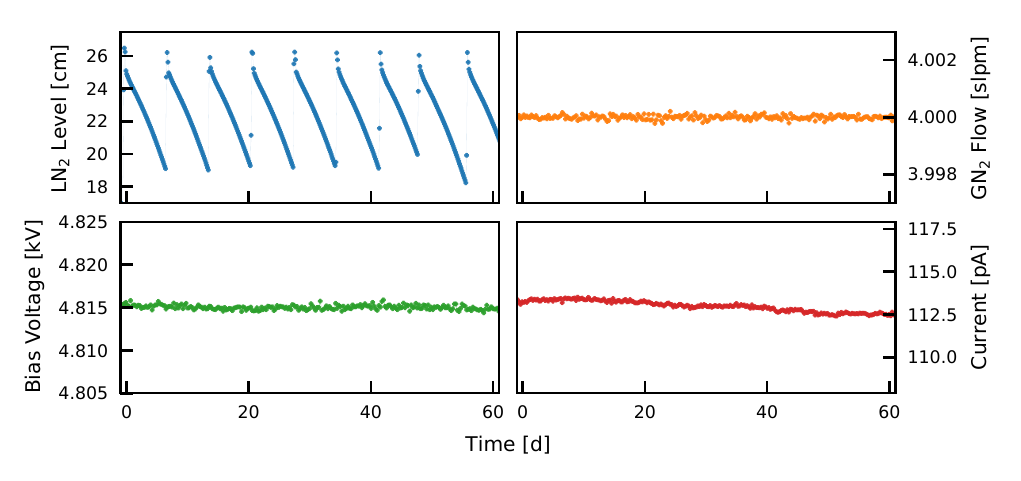}
\caption{Time evolution of selected Gator operational parameters: the $\mathrm{LN}_2$ level in the dewar, the $\mathrm{GN}_2$ purge flow, as well as the detector bias voltage and leakage current. Data points, shown in 6-hour bins, are from a 2-month period towards the end of 2021. In the top left plot, the weekly refill cycles of the $\mathrm{LN}_2$ dewar, which is connected to the coldfinger of the detector, are clearly discernible as a sharp rise during each refill, followed by a continuous decrease in level as the $\mathrm{LN}_2$ evaporates. The fluctuations of the $\mathrm{GN}_2$ purge flow, detector bias voltage and leakage current are at the sub-percent level and do not affect the energy calibration.}
\label{fig:sc_stability}
\end{figure}

The stability of the high voltage, the leakage current, $\mathrm{LN}_2$ refill cycles and $\mathrm{GN}_2$ flow in a period of 2 months in late 2021 are shown in \autoref{fig:sc_stability} as an example. The values of the monitored parameters can also be used to correlate operations on the detector or environmental changes with artefacts in the data. For example, we found an increase in low-energetic noise, up to \SI{150}{keV}, during refills of the $\mathrm{LN}_2$ dewar. This information is employed for an unbiased removal of these periods from the data, leading to a live-time reduction of about \SI{2}{\percent}. 

\begin{figure}[htp]
\centering
\includegraphics[width=1\textwidth]{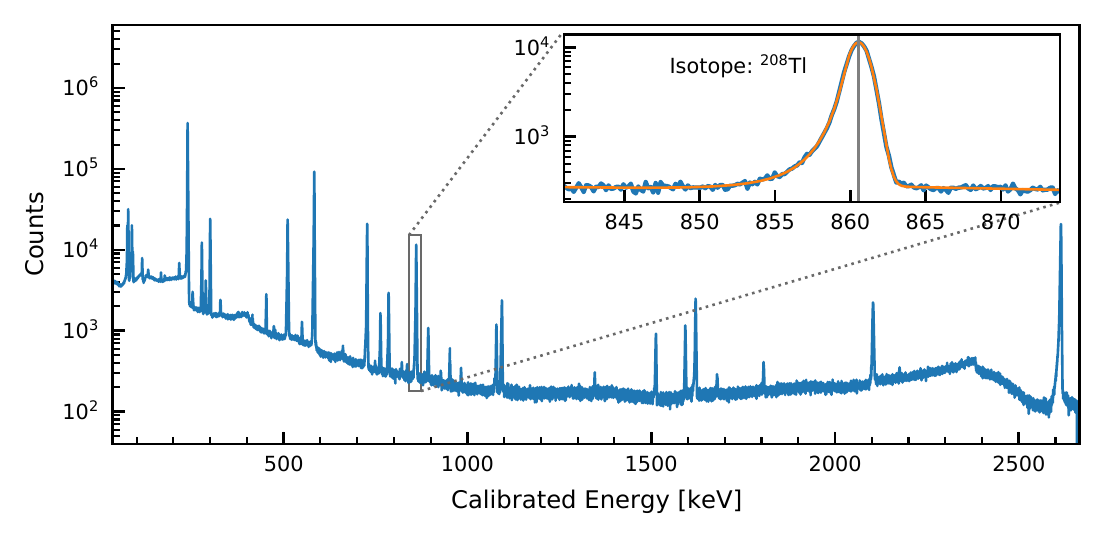}
\caption{A $^{228}$Th spectrum calibrated with the known positions of its peaks. 
The inset shows a fit of the \SI{860.5}{keV} $^{208}$Tl line with a combination of a Crystal Ball function, a linear background as well as a smeared step function.}
\label{fig:th_calibration}
\end{figure}

Regular calibrations of the detector with radioactive sources such as \mbox{$^{228}$Th}, \mbox{$^{137}$Cs}, or \mbox{$^{60}$Co}, as well as with certified extended sources and composite, high-activity materials, ensure proper knowledge of the detected energy and its resolution. For this purpose, $\gamma$-ray peaks of selected isotopes are modeled with a combination of a Crystal Ball function (a Gaussian core part for the peak with a low-end power-law tail to model loss effects~\cite{Skwarnicki:1986xj}), a linear background for the continuum, and a smeared step function for background discontinuities at Compton edges. An exemplary $^{228}$Th calibration spectrum is shown in \autoref{fig:th_calibration}. 

The determined peak positions are fitted to their literature values with a polynomial of order one. The energy-dependent resolution is described by a quadratic function for the variance, resulting in a FWHM at \SI{1332}{keV} of \SI{2.03(4)}{keV}. This value is shown in \autoref{comparison_table}, along with the energy resolution of some of the most sensitive HPGe detectors for $\gamma$-ray spectrometry.

\begin{table}[htp]
    \centering
    \caption{Location, mass, efficiency, energy resolution (FWHM at \SI{1332}{\keV}) and background rate of Gator in comparison to other $\gamma$-ray screening HPGe detectors located in SURF~\citep{MOUNT2017130}, LNGS, Boulby~\citep{SCOVELL2018160}, LSC~\citep{LSC_det}, and LVdA~\citep{doi:10.1063/1.1614854}. Not all references provide the uncertainty in the count rate.
    }
    \vspace{0.3cm}

	\begin{tabularx}{0.93\textwidth}{lccccc}
\hline
Detector & Location & Crystal & Efficiency & FWHM at & Rate \\
 & (Overburden & Mass & & \SI{1332}{\keV} & 60-2700 keV  \\
 & [m.w.e.]) & [kg] & & [keV] &  [counts/(kg$\cdot$day)] \\
\hline
\textbf{Gator} & LNGS (3600) & 2.2 & 100.5 \% & 1.98 & \SI{89.0(7)}{} \\
\hline
\textbf{Maeve}~\citep{LZ:2020fty} & SURF (4300) & 2.0 & \phantom{1}85 \%\phantom{1.} & 3.19 & 956.1 \\
\textbf{GeMPI 3}~\citep{priv_comm} & LNGS (3600) & 2.2 & \phantom{1}98.7 \% & 2.20 & \SI{24(1)}{}  \\
\textbf{Belmont}~\citep{LZ:2020fty} & Boulby (2805) & 3.2 & 160 \%\phantom{1.} & 1.92 & 135.0  \\
\textbf{GeOroel}~\citep{LSC_det} & LSC (2450) & 2.2 & 109 \%\phantom{1.} & 1.85 & 165.3  \\
\textbf{GeMSE}~\cite{Garcia:2022jdt, Diego_com} & LVdA (620) & 2.0 &  107.7 \% & 1.96 & \SI{88(1)}{}   \\
\hline
\end{tabularx}
    \label{comparison_table}
\end{table}

\section{Background analysis}
\label{sec:background}

\begin{figure}[h!]
\centering
\includegraphics[width=1\textwidth]{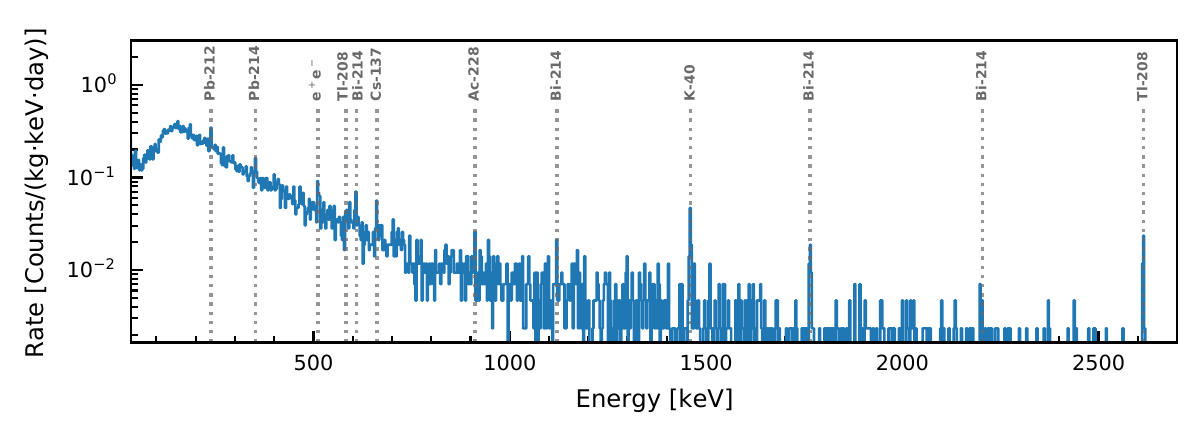}
\caption{Recent Gator background spectrum (October -- December 2021) using \SI{73.7}{days} of data. Prominent $\gamma$-ray lines are labeled.}
\label{fig:background_comparison}
\end{figure}

After the detector upgrade and the cleaning of all shielding components, we have acquired \SI{74}{days} of background data, shown in \autoref{fig:background_comparison}. In the energy region between \SIrange{35}{2700}{keV}, the background is dominated by $\gamma$-rays from detector and shielding materials, primarily from the primordial $^{238}$U, $^{232}$Th, $^{235}$U and $^{40}$K. Of particular concern is the gaseous $^{222}$Rn daughter from primordial $^{238}$U, which is present in the laboratory environment. Although a dedicated ventilation system in the underground laboratory helps to maintain a relatively low level of radon in the experimental facilities~\cite{bassignani1995review}, one of the motivations for the upgrade, as described in \autoref{sec:upgrade}, was to tighten the enclosure so that the $\mathrm{GN}_2$ purge displaces the environmental radon more effectively.

The integrated background rate is $\SI{82.0(7)}{counts/(kg\cdot day)}$ in the region \SIrange{100}{2700}{keV}, as compared to the value of $\SI{102.8(7)}{counts/(kg\cdot day)}$ in 2010~\citep{Baudis:2011am}. We quote the rates for the region $60-2700$ keV in \autoref{comparison_table} to allow for comparison with some of the most sensitive screening facilities. Most of these present low count rates of $\mathcal{O}(100)$ $\si{counts/(kg\cdot day)}$, with the lowest count rate being for GeMPI~3~\citep{priv_comm}, followed by  GeMSE~\cite{Garcia:2022jdt} and Gator. Also given in \autoref{comparison_table} are the detector locations and overburden in meter water equivalent (m.w.e), the mass of each HPGe crystal, along with the detector efficiencies and FWHM at $\SI{1332}{keV}$.

\begin{table} [htpb]
		\centering
		\caption{Compton-subtracted rates in the 3$\sigma$-regions for selected $\gamma$-ray lines in the background spectrum. The rates of 3 different runs between 2010 and 2021 are shown, with live-times in days of 73.7 (2021), 98.5 (2015), and 51.4 (2010), respectively.}
			\begin{tabularx}{0.85\textwidth}{cc|ccc}
				\hline
				\multicolumn{2}{c|}{} & \multicolumn{3}{c}{Rate [counts/day]}\\
				Energy [keV] & Chain/nuclide & October 2021 & October 2015 & April 2010~\citep{Baudis:2011am} \\ 
				\hline
				351.932 & $^{238}$U/$^{214}$Pb & 0.41 $\pm$ 0.17 & 0.58 $\pm$ 0.14 & 0.7 $\pm$ 0.3 \\
				609.312 & $^{238}$U/$^{214}$Bi & 0.26 $\pm$ 0.10 & 0.48 $\pm$ 0.09 & 0.6 $\pm$ 0.2 \\
				1120.29 & $^{238}$U/$^{214}$Bi & < 0.28 & 0.18 $\pm$ 0.06 & 0.3 $\pm$ 0.1 \\
				1764.49 & $^{238}$U/$^{214}$Bi & 0.14 $\pm$ 0.06 & 0.18 $\pm$ 0.05 & 0.08 $\pm$ 0.06 \\
				661.657 & $^{137}$Cs & 0.19 $\pm$ 0.09 & 0.24 $\pm$ 0.07 & 0.3 $\pm$ 0.1 \\
				1173.24 & $^{60}$Co & < 0.27 & < 0.34 & 0.5 $\pm$ 0.1 \\
				1332.51 & $^{60}$Co & < 0.21 & < 0.24 & 0.5 $\pm$ 0.1 \\
				1460.88 & $^{40}$K & 0.28 $\pm$ 0.08 & 0.43 $\pm$ 0.07 & 0.5 $\pm$ 0.1 \\
				2614.51 & $^{232}$Th/$^{208}$Tl & 0.19 $\pm$ 0.05 & 0.22 $\pm$ 0.05 & 0.2 $\pm$ 0.1 \\
				\hline
			\end{tabularx}
		\label{tab:LineActivity}		
	\end{table}

\autoref{tab:LineActivity} presents the integral count rates of $\gamma$-ray lines which are typically present in the background data -- these are lines from primordial isotopes (or from their daughters) as well as from medium-lived cosmogenic or anthropogenic isotopes which have a high branching ratio or detection efficiency. The count rates of these $\gamma$-ray lines in the new background data are consistent within the uncertainties or lower than the values measured in 2010~\citep{Baudis:2011am}. 
As expected, the activities of cosmogenic radionuclides with half-lives on the order of a few years, such as $^{60}$Co, have noticeably decreased while underground. Although still consistent within errors, the activities of $^{40}$K and isotopes from the $^{222}$Rn chain (e.g. $^{214}$Pb, $^{214}$Bi) seem to have decreased as well, likely due to the cleaning of the detector cavity and/or tightening of the enclosure.

\section{Sample analysis and results}
\label{sec:samples}

Before a sample is submitted to a radioassay in Gator, we  evaluate whether its mass, material, and expected radioactivity are likely to yield either values within the sensitivity reach (after approximately \SI{30}{days} of screening) or upper limits that fulfill the radiopurity requirements for the given sample. This evaluation is performed by comparing the sample specifications to samples previously measured in Gator and their achieved sensitivity. Once the sample has been selected and measured in Gator, its data analysis is performed following a counting method around the $\gamma$-ray lines of interest. This section summarises this analysis procedure and presents an example of a radioassay of photosensors conducted with the new background level.

\subsection*{Analysis procedure}

To estimate the activity of the samples, we perform a Geant4 simulation~\citep{GEANT4:2002zbu} using a model of the exact geometry, chemical composition and density of materials present in the sample, detector and cavity. 
The simulation of the detector geometry and its components are validated using calibrated point and extended sources, as described in detail in~\citep{Baudis:2011am}. 
A source term for each expected isotope or decay chain is distributed uniformly throughout the sample. The decays and their subsequent interactions within the modeled geometry are simulated via a Monte Carlo method in Geant4. The final energy depositions within the detector are then stored and used to calculate the efficiency of each gamma line. The number of simulated decays in the sample volume is typically of $\mathcal{O}(10^{8})$ for each isotope of interest. The ratio between the detected and the simulated number of gammas results in the detection efficiency ($\varepsilon$). This efficiency is geometry-, material-, and energy-dependent, as will be shown with the photosensor sample. 

Background runs are acquired in between the materials screening, not longer than 12 months before or after the sample measurement, with acquisition times ($t_{B}$) usually longer than that of the sample ($t_{S}$). 
Although the samples are usually pre-purged in the load-lock chamber, the first hours/days of data acquisition may present a higher rate due to an initially larger concentration of $^{222}$Rn, which has a half-life of \SI{3.82}{days} and is still being purged out of the cavity or emanating from the samples. The trigger rate can be thus modeled with an exponential function during this initial period and later reaches a constant level within uncertainties. This initial period, for which the rate is dominated by an exponential behavior, as well as data in the direct vicinity of $\mathrm{LN}_2$ refills (described in \autoref{sec:operation}), are thus removed from the analysis data set. 
The acquired sample and background data are then calibrated using the functions obtained with dedicated calibration runs. 

After the data selection and calibration, we count the number of events in a $3\sigma$ region around the centroid of $\gamma$-ray lines of interest in both the sample and the background data. The Compton background is estimated from the left and right sidebands around each peak. These values are then used to obtain the Compton-subtracted counts within each $\gamma$-ray peak in the sample (\textit{S}) and in the background (\textit{B}) data. The latter is corrected for the acquisition time in order to obtain the net signal $S_{net}$ from the sample for a given $\gamma$-ray line:

\begin{equation}\label{eq:snet}
    S_{net}=S-B\cdot \frac{t_{S}}{t_{B}}. 
\end{equation}

The background and sample rates are usually low (with both \textit{S} and \textit{B} below 100 counts in the peak, as exemplified in \autoref{tab:LineActivity}); we thus use the method of activities and upper limit determination for low-radioactivity measurements detailed in~\citep{HURTGEN200045}. We estimate the standard uncertainties on $x$ counts as $\sqrt{x+1}$ and implement these in the calculation of the detection limit $L_{d}$ which is defined as the level of true net signal that leads to a certain probability of detection~\citep{HURTGEN200045}. Following the recommendation of the same reference, we report activity values if $S_{net}$ exceeds $L_{d}$, and report upper limits on the activity otherwise. 
To translate $S_{net}$ from a given $\gamma$-ray line to the specific activity of the isotope, we use the following equation: 

\begin{equation}\label{eq:spec_activ}
    A[\mathrm{Bq/kg}] = \frac{S_{net}}{r \cdot \varepsilon \cdot m \cdot t_{S}},
\end{equation}

\noindent where \textit{r} is the branching ratio of the $\gamma$-ray line and \textit{m} is the sample mass. To calculate the error on the activity, we take both the error on $S_{net}$ and on the efficiency into account, with the latter being a \SI{10}{\percent} systematic error (described in~\citep{Baudis:2011am}). If several lines of a given isotope are present in the spectrum, with $S_{net}>L_{d}$, the activity is calculated as the error-weighted average.

\subsection*{Photosensor sample}
\label{sec:PMT}
After the upgrade, we have screened two Hamamatsu R12699-406-M4 photomultiplier tubes (PMTs)~\cite{Hamamatsu:R12699}, which are a viable candidate photosensor for the future DARWIN detector. The baseline design of DARWIN features a liquid-xenon target instrumented by two large arrays of photosensors~\cite{Aalbers:2016jon}. The sample photosensor has a square 2-inch photosensitive area, in contrast to the circular 3-inch R11410 units used in XENON1T~\cite{Aprile:2015lha}, XENONnT~\citep{Antochi:2021wik}, LZ~\cite{Mount:2017qzi} and PandaX-4T~\cite{PandaX-4T:2021bab}. The square shape allows for a higher packing density in an array as compared to circular tubes. Together with a higher photocathode coverage of \SI{75.0}{\percent}, as opposed to \SI{61.8}{\percent} for R11410 PMTs at densest packing, this implies an increase in the light collection efficiency of the detector. Furthermore, the 2-inch sensor possesses a buoyancy that is two orders of magnitude smaller in xenon. This feature suggests that an array of R12699 PMTs would require a less rigid mechanical support structure and hence less material close to the xenon target. As such, it would be feasible to reduce the backgrounds that arise from detector materials if, in addition, the average radioactivity of the R12699 unit can be reduced sufficiently.

\begin{figure}
	\centering
	\begin{subfigure}[t]{0.45\textwidth}
		\includegraphics[width=\textwidth]{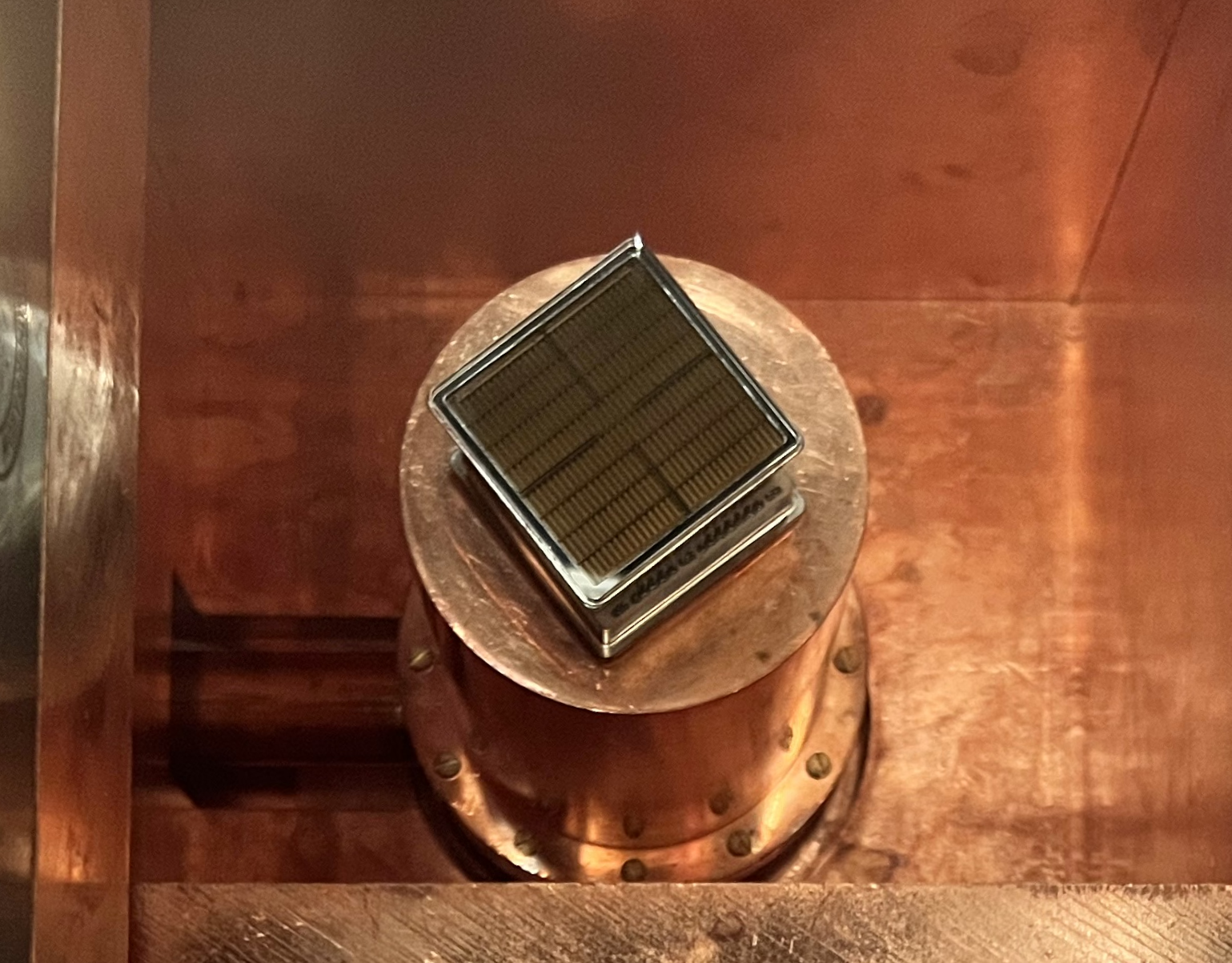}
	\end{subfigure}
	~ 
	\begin{subfigure}[t]{0.45\textwidth}
		\includegraphics[width=\textwidth]{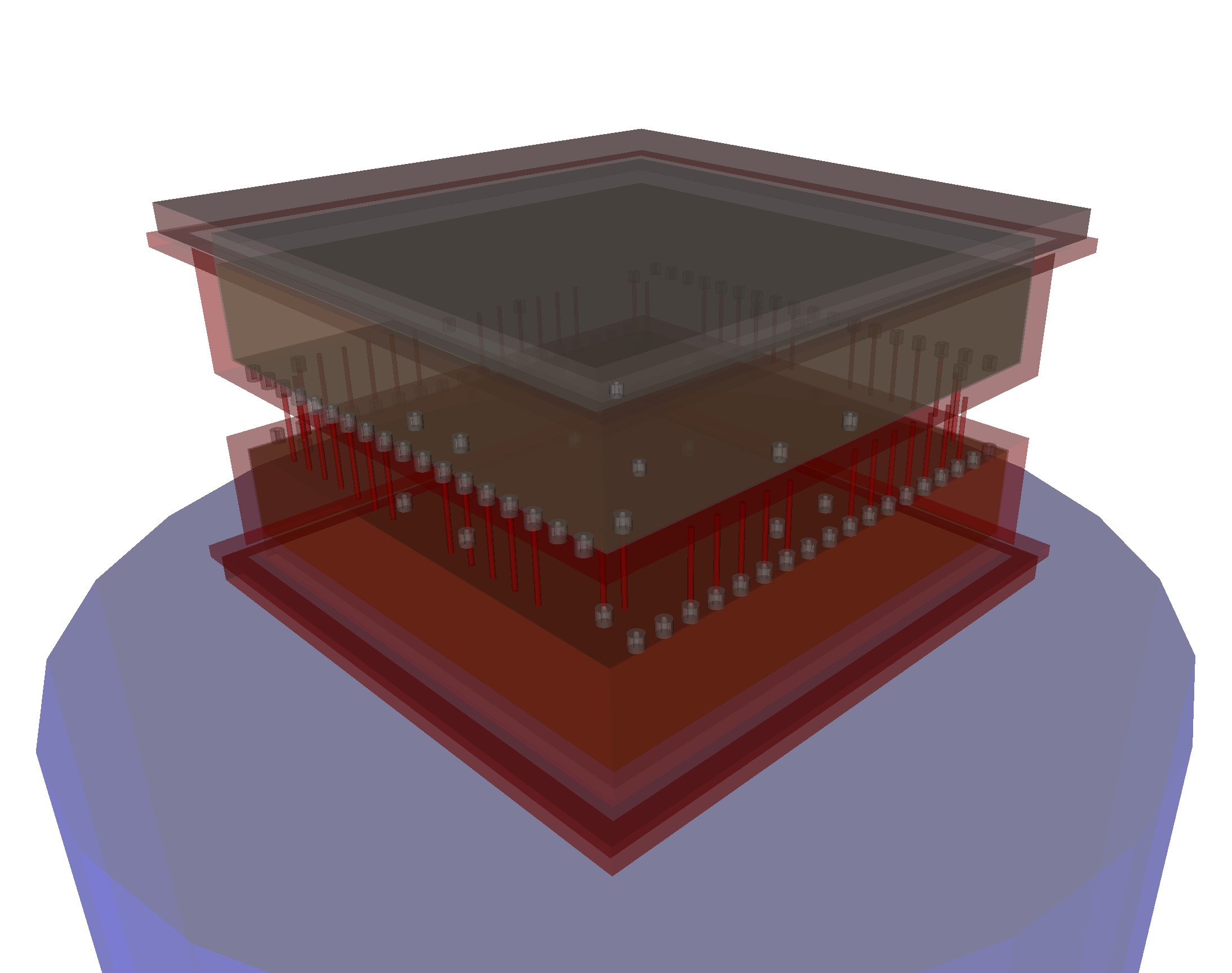}
	\end{subfigure}
	\caption{(Left) Two Hamamatsu R12699-406-M4 PMTs inside the Gator sample cavity. (Right) Geometry as implemented in the simulations. The upper part of the copper cryostat around the HPGe crystal is shown in blue. For the PMTs, the metal envelope and pins are colored red, the inner effective dynode material orange, and the silica window together with the  glass around the pins, white. All other simulated components, such as the sample cavity, are not shown for clarity.}\label{fig:geom}
\end{figure}

\begin{figure}
	\centering
	\includegraphics[width=1.\textwidth]{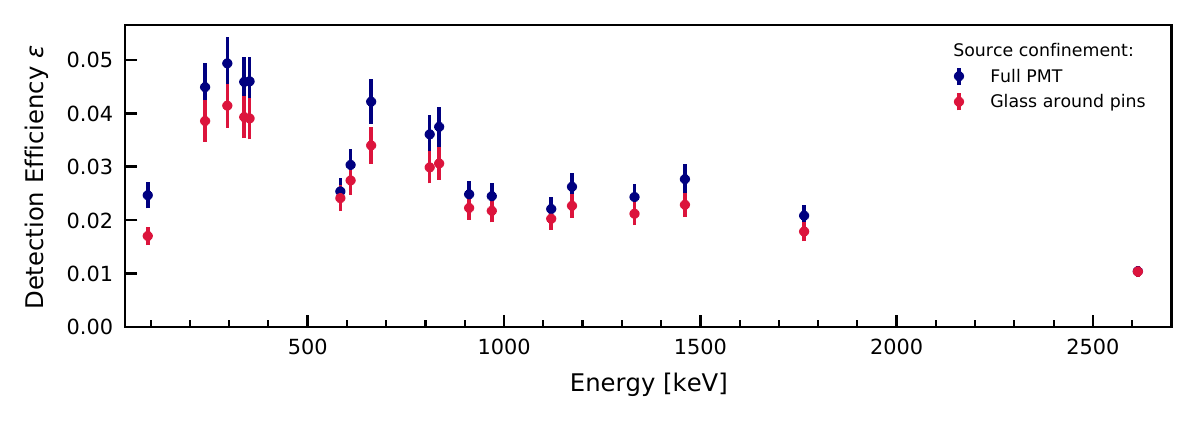}
	\caption{Simulated detection efficiencies for selected lines in the sample PMTs described in \autoref{sec:PMT}. Values for a confinement of the impurities in the simulations to the full PMT and only the glass around the pins, respectively, are shown. The latter yields a systematically lower detection efficiency due to geometrical and material absorption effects.}
	\label{fig:efficiencies}
\end{figure}

\autoref{fig:geom} shows the PMTs on the top of the detector as well as the geometry implemented in the simulations of the detection efficiencies. The resulting efficiencies for selected $\gamma$-ray lines are depicted in \autoref{fig:efficiencies}. Below approximately \SI{300}{keV}, the detection efficiency increases with energy as more gammas are able to penetrate the sample itself as well as the copper cryostat and dead layer of the HPGe crystal. At higher energies, it decreases again as energetic gammas will not fully deposit their energy within the sensitive crystal volume. The finite time resolution of the detector may lead to the registration of two temporally unresolvable $\gamma$-ray depositions as a single summation peak. Gammas resulting in a summation peak are hence missing in the peaks of their individual energies which are consequently systematically reduced. 
For example, the reduced efficiency at \SI{583.2}{keV} is due to a summation peak at \SI{3197.7}{keV} from the \SI{583.2}{keV} and \SI{2614.5}{keV} $\gamma$-ray lines of $^{208}$Tl.

To demonstrate the impact of the assumptions in the efficiency simulations, especially for complex geometries, we compare two different spatial distributions of the decays: a homogeneous source confinement to the entire PMT and an assumed localization of the radioimpurities to only the glass around the pins (visible as white cylinders in \autoref{fig:geom}). A lower detection efficiency is obtained for the latter, as shown in \autoref{fig:efficiencies}, due to geometrical and material absorption effects. This in turn would lead to higher derived activities following \autoref{eq:spec_activ}. Thus the spatial distribution of radioimpurities is one of the factors included in the systematic uncertainty of the final activity values.

\begin{figure}[htp]
    \centering
    \includegraphics[width=1.\textwidth]{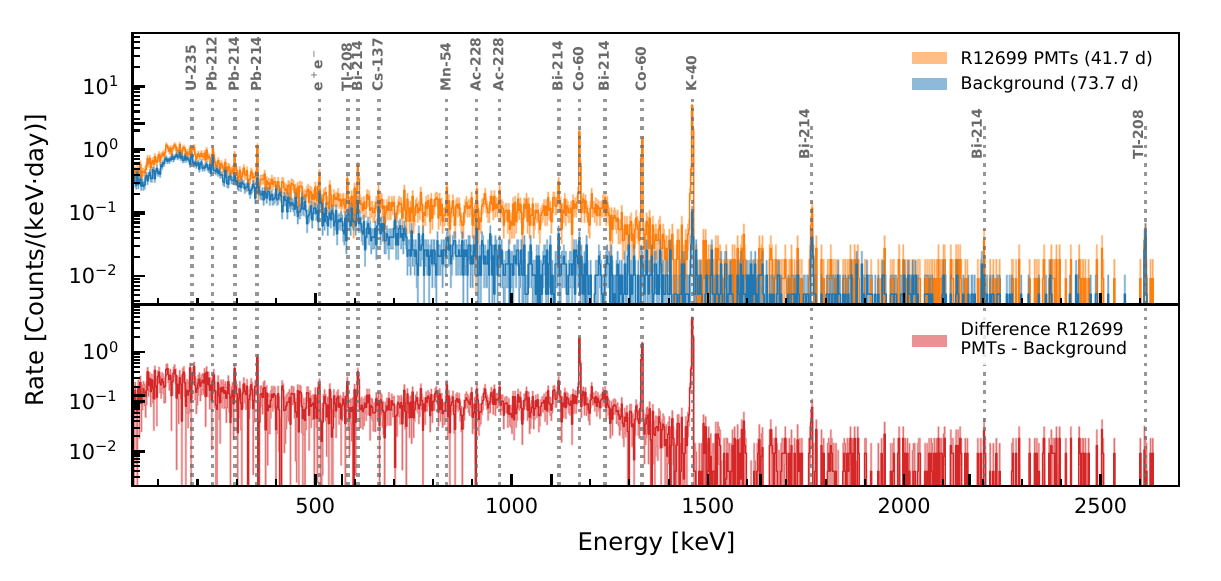}
    \caption{Measured energy spectrum of the two Hamamatsu R12699 PMTs (orange) compared to the background spectrum (blue). The background-subtracted spectrum is given in red. Statistical uncertainties are indicated as transparent bands. Prominent $\gamma$-ray lines from various isotopes are labeled.}
    \label{fig:pmtspectrum}
\end{figure}

The analysis of this sample is based on $\sim$$\SI{41.7}{days}$ of data from the screening of the two PMTs together with the $\sim$$\SI{73.7}{days}$ of the background reported in \autoref{sec:background}. The spectra of these two data sets are shown in \autoref{fig:pmtspectrum}. Given in \autoref{tab:ActivitiesSummaryPMTs} are the resulting specific activities as compared to Gator screening results for the Hamamatsu R11410-21 PMTs employed in XENON1T and XENONnT~\cite{Aprile:2017ilq} as well as the Hamamatsu R8520-06 PMTs used in XENON100~\cite{Aprile:2011ru}. These values are all based on detection efficiency simulations assuming a source confinement to the entire PMT for consistent comparison. 
For the isotopes $^{228}$Th, $^{60}$Co and $^{40}$K, the activity per active photocathode area of the current R12699 model is lower compared to the R8520 PMTs~\cite{Aprile:2011ru}. For the other isotopes listed in \autoref{tab:ActivitiesSummaryPMTs}, the comparison is inconclusive, as only upper limits are given for their values (particularly in the measurement of the R8520 PMTs).   With respect to the R11410-21 units~\cite{Aprile:2017ilq}, the sample PMT has from equal to fourfold higher activities per active photocathode area. We note, however, that the R12699 PMT is currently being optimized in terms of low-activity materials. The eventually achievable activity, together with further parameters, such as the performance of the PMT which is currently being characterized, will decide on its applicability in DARWIN.

	\begin{table}
			\caption{Summary of activities for the main isotopes from the two Hamamatsu R12699 PMTs. Shown for comparison are the Gator results for the Hamamatsu R11410-21 PMTs used in XENON1T and XENONnT~\cite{Aprile:2017ilq} and the Hamamatsu R8520-06 PMTs employed in XENON100~\cite{Aprile:2011ru}. The activities are given per PMT (top) and per active photocathode area (bottom).
			}
		\centering
		\small{
			\begin{tabularx}{0.58\textwidth}{cccc}
				\hline
				Isotope & R12699-406 & R11410-21 & R8520-06\\
				\hline
				\hline
				\multicolumn{4}{c}{Activity [mBq/PMT]}\\
				\hline
				$^{238}$U & < 6.1 & 8 $\pm$ 2 & < 15 \\
				$^{226}$Ra & 0.60 $\pm$ 0.10 & 0.6 $\pm$ 0.1 & < 0.28 \\
				$^{228}$Ra & < 0.65 & 0.7 $\pm$ 0.2 & < 0.59 \\
				$^{228}$Th & < 0.53 & 0.6 $\pm$ 0.1 & 0.3 $\pm$ 0.1 \\
				$^{235}$U & < 0.28 & 0.37 $\pm$ 0.09 & < 0.67 \\
				$^{60}$Co & 1.31 $\pm$ 0.11 & 0.84 $\pm$ 0.09 & 0.60 $\pm$ 0.04 \\
				$^{40}$K & 35 $\pm$ 4 & 12 $\pm$ 2 & 12.0 $\pm$ 0.8 \\
				$^{137}$Cs & < 0.12 & -- & < 0.10 \\
				\hline
				\hline
				\multicolumn{4}{c}{Activity [mBq/cm$^2$]}\\ 
				\hline
				$^{238}$U & < 0.260 & 0.25 $\pm$ 0.06 & < 3.569 \\
				$^{226}$Ra & 0.026 $\pm$ 0.004 & 0.019 $\pm$ 0.003 
				& < 0.067\\
				$^{228}$Ra & < 0.028 & 0.022 $\pm$ 0.006 & < 0.140 \\
				$^{228}$Th & < 0.023 & 0.019 $\pm$ 0.003  & 0.071 $\pm$ 0.017 \\
				$^{235}$U & < 0.012 & 0.012 $\pm$ 0.003 & < 0.159 \\
				$^{60}$Co & 0.055 $\pm$ 0.005 & 0.026 $\pm$ 0.003 & 0.144 $\pm$ 0.010 \\
				$^{40}$K & 1.47 $\pm$ 0.16 & 0.37 $\pm$ 0.06 & 2.86 $\pm$ 0.18 \\
				$^{137}$Cs & < 0.005 & -- & < 0.024 \\
				\hline
			\end{tabularx}
		}
		\label{tab:ActivitiesSummaryPMTs}
	\end{table}

This exemplary sample also demonstrates the sensitivity gain from the reduced background rate. For instance, for the given background data, the counts around the $^{214}$Bi line at \SI{609.3}{keV} exceed the detection limit after a sample measurement time of about \SI{20}{days}. At an overall \SI{25}{\percent} higher background (similar to the integrated background rate in 2010~\citep{Baudis:2011am}), additional sample data taking of $\sim$$\SI{15}{\percent}$ would be necessary. For lines with longer required measurement times (due to a larger background at their energy, lower detection efficiency or branching ratio, or lower activity), this effect is even greater.

\section{Summary and conclusions}
\label{sec:summary}

In this work, we described an upgrade of the Gator HPGe $\gamma$-ray spectrometry facility with respect to the detector enclosure and sample-handling infrastructure. Following the improvements, a background rate of \SI{82.0(7)}{counts/(kg\cdot day)} in the \SIrange{100}{2700}{keV} region was achieved over a 74-day counting period, reflecting a \SI{20}{\percent} decrease with respect to the previously reported value~\cite{Baudis:2011am}. 
This reduction can be attributed to the reduced activity of the cosmogenic isotope $^{60}$Co over time ($\mathrm{T}_{1/2} = \SI{5.3}{yrs}$) and also likely due to the cleaning of the detector cavity and tightening of the enclosure. Further, the new enclosure improved the handling of samples, reducing the risk of introducing contamination into the cavity.

The HPGe detector has been under stable operation for over a decade, ensured by constant monitoring of its operating parameters as well as the facility environment. It undergoes regular calibrations, which allow us to monitor the energy reconstruction, resolution, and efficiency. Recent calibrations yield an energy resolution of \SI{2.03(4)}{keV} at \SI{1332}{keV} (FWHM).

We summarised the analysis procedure, optimised for ultra-low radioactivity samples. This analysis relies on the detection efficiency for each specific radioisotope and sample, estimated by means of Geant4 simulations of the precise detector and sample geometries. As an example from the ongoing DARWIN radioassay campaign, we report results from a 42-day screening of the Hamamatsu R12699-406-M4 PMT, a possible alternative to the photosensors currently used by some of the most sensitive dark matter direct detectors. 

Gator was previously used for the XENON dark matter and GERDA, LEGEND-200 neutrinoless double beta decay experiments and will be used in the future for LEGEND-1000 and for DARWIN. These experiments target yet lower backgrounds and hence more radiopure materials, which in turn requires even higher sensitivities in their radioassay, that could be achieved in Gator thanks to the reduced background rate.

\acknowledgments
This work was supported by the Swiss National Science Foundation under Grant No. 200020-188716, by the Candoc Grant No. K-72312-09-01 from the University of Zurich, by the European Unions Horizon 2020 research and innovation programme under the Marie Sklodowska-Curie grant agreements No. 690575 and No. 674896, and  by the European Research Council (ERC) under the European Union's Horizon 2020 research and innovation programme, grant agreement No. 742789 ({\sl Xenoscope}). We thank Andreas James, Francesco Piastra, Roberto Corrieri, and Junji Naganoma for their valuable contributions to the facility upgrade as well as Diego Ram\'\i{}rez Garcia for his careful review of this manuscript. We also thank the mechanical workshop in the Physics Department for their continuous support and the LNGS mechanical and electrical technicians for the support provided onsite.

\bibliographystyle{JHEP}
\bibliography{gator_upgrade} 

\end{document}